\documentclass[conference]{IEEEtran}
\IEEEoverridecommandlockouts
\usepackage{cite}
\usepackage{amsmath,amssymb,amsfonts}
\usepackage{algorithmic}
\usepackage{graphicx}
\usepackage{physics}
\usepackage{comment}
\usepackage{textcomp}
\usepackage{xcolor}
\usepackage{booktabs}
\usepackage{subcaption}
\usepackage{hyperref}

\usepackage[capitalize]{cleveref}
\crefname{equation}{Eq.}{Eqs.}
\Crefname{equation}{Eq.}{Eqs.}
\crefname{subfigure}{Fig.}{Figs.}
\Crefname{subfigure}{Fig.}{Figs.}

\def\BibTeX{{\rm B\kern-.05em{\sc i\kern-.025em b}\kern-.08em
    T\kern-.1667em\lower.7ex\hbox{E}\kern-.125emX}}
\begin{document}

\title{
Adaptive Clifford+T Decomposition of Large Toffoli Gates with One Clean Ancilla
%Adaptive Clifford+T Decomposition of MCT Gates with One Clean Ancilla
%Late Breaking Results: Dynamic Clean-Ancilla Approach for Efficient 5-Input Toffoli Realization\\
%Late Breaking Results:Dynamic Clean-Ancilla Approach for Low-Cost Realization of the 5-Input Toffoli Gate\\
}
\author{
Abhoy~Kole\textsuperscript{$\Delta$},
Majd~Assaad\textsuperscript{$\Delta$}
Till~Schnittka\textsuperscript{$\Delta$},
Rolf~Drechsler\textsuperscript{$\Delta$,$\Cup$}\\%~\IEEEmembership{Fellow,~IEEE}\\
 
 \textsuperscript{$\Delta$}\textit{Cyber-Physical Systems, DFKI, Bremen, Germany} \\ 
 \textsuperscript{$\Cup$}\textit{Institute of Computer Science, University of Bremen, Bremen, Germany}\\  
 \{abhoy.kole, majd.assaad, till.schnittka\}@dfki.de, drechsler@uni-bremen.de }

\title{Adaptive Clifford+T Decomposition of Large Toffoli Gates with One Clean Ancilla\\
%{\footnotesize \textsuperscript{*}Note: Sub-titles are not captured for https://ieeexplore.ieee.org  and should not be used}
\thanks{This work was partly funded by the Federal Ministry of Research, Technology and Space (BMFTR) through the EASEPROFIT project (grant no.~16KIS2127) and the EQeS project (grant no.~13N17378), and by the German Research Foundation (DFG) through the CONAD-QC project (grant no.~559888852). This research was conducted within the scope of the DFG Priority Programme~2514: Quantum Software, Algorithms and Systems.}
%This work was partly funded by the Federal Ministry of Research, Technology and Space (BMFTR) through the EASEPROFIT project (grant no.~16KIS2127) and the EQeS project (grant no.~13N17378), and by the German Research Foundation (DFG) through the CONAD-QC project (grant no.~559888852). This research was conducted within the scope of the DFG Priority Programme~2514: Quantum Software, Algorithms and Systems (SPP~2514).
}

\author{\IEEEauthorblockN{Abhoy~Kole}
\IEEEauthorblockA{\textit{Cyber-Physical Systems} \\
\textit{DFKI GmbH}\\
Bremen, Germany \\
abhoy.kole@dfki.de}
\and
\IEEEauthorblockN{Majd~Assaad}
\IEEEauthorblockA{\textit{Cyber-Physical Systems} \\
\textit{DFKI GmbH}\\
Bremen, Germany \\
majd.assaad@dfki.de}
\and
\IEEEauthorblockN{Till~Schnittka}
\IEEEauthorblockA{\textit{Cyber-Physical Systems} \\
\textit{DFKI GmbH}\\
Bremen, Germany \\
till.schnittka@dfki.de}
\and
\IEEEauthorblockN{Rolf~Drechsler}
\IEEEauthorblockA{\textit{Institute of Computer Science} \\
\textit{University of Bremen / DFKI GmbH}\\
Bremen, Germany \\
drechsler@uni-bremen.de}
}

%\author{
%Anonymous Authors \\%~\IEEEmembership{Fellow,~IEEE}\\
%  \\ 
% \\  
% \\}

\maketitle

\begin{abstract}
Multi-controlled Toffoli gates are fundamental building blocks in quantum computation, with applications in quantum arithmetic, simulation, and search algorithms. In fault-tolerant architectures, their realization is constrained by the high cost of non-Clifford resources, particularly in terms of T-count and T-depth. Recent advances have demonstrated that the use of ancillary qubits, relative-phase Toffoli gates, and dynamic circuit techniques can substantially reduce this overhead.

In this work, we investigate the decomposition of large Toffoli gates using 3- and 4-input relative-phase Toffoli gates in the presence of a single clean ancilla and conditionally clean ancillas. 
We derive explicit resource bounds for Clifford+T implementations incorporating dynamic-circuit-based uncomputation and measurement-conditioned corrections. Our analysis emphasizes T-depth reduction under fixed CX and T-count overhead, ensuring relevance for near-term devices. We show that introducing 4-input relative-phase Toffoli gates enables significant T-depth reductions through enhanced parallelism while maintaining favorable ancilla requirements. We further validate our theoretical results through experimental evaluation and comparative analysis with existing approaches. 
\end{abstract}


\section{Introduction}

%The Toffoli gate and its multi-controlled extensions constitute fundamental primitives in quantum computation, with applications spanning diverse domains such as quantum arithmetic~\cite{9655478}, quantum simulation~\cite{Rocca2024}, and the implementation of oracles for Grover’s algorithm~\cite{10.1145/3729229}. Despite their importance, the efficient synthesis of large Toffoli gates remains a central challenge. Standard Clifford+T decompositions typically require a large number of T gates, incur substantial circuit depth, and rely heavily on CX-based entangling structures~\cite{PhysRevA.93.022311}.  These constraints are particularly restrictive for Noisy Intermediate-Scale Quantum (NISQ) devices, where noise, decoherence, and limited qubit availability severely constrain the achievable circuit depth and circuit width. In fault-tolerant architectures, Toffoli gates are realized using magic-state-based constructions. Consequently, their implementation cost is directly tied to T-gate resources, making both T-count and T-depth major performance bottlenecks~\cite{Litinski2019gameofsurfacecodes}.

The Toffoli gate and its multi-controlled extensions are fundamental primitives in quantum computation, with applications in quantum arithmetic~\cite{9655478}, quantum simulation~\cite{Rocca2024}, and Grover-oracle implementations~\cite{10.1145/3729229}. However, efficient synthesis of large Toffoli gates remains challenging, as standard Clifford+T decompositions require high T-count, large circuit depth, and extensive CX-based entangling structures~\cite{PhysRevA.93.022311}. These overheads are particularly restrictive for Noisy Intermediate-Scale Quantum (NISQ) devices due to noise, decoherence, and limited qubit resources. In fault-tolerant architectures, Toffoli gates are typically realized via magic-state-based constructions, making both T-count and T-depth critical cost metrics~\cite{Litinski2019gameofsurfacecodes}.

Designing resource-efficient circuits for large Toffoli gates with minimal ancilla overhead therefore remains an active area of research. In many constructions, reductions in T-gate resources can be achieved by introducing additional ancilla qubits~\cite{PhysRevA.111.052611,10.1007/978-3-031-08760-8_16}. In particular, the use of conditionally clean ancillas has been shown to be especially beneficial for realizing large Toffoli gates using fewer non-Clifford operations~\cite{Khattar2025riseofconditionally,nie2024quantum}. Such techniques lead to improved resource bounds when one or two clean ancillas are available.

Recent advances in dynamic quantum circuits, which permit intermediate measurements, qubit resets, and classically controlled operations, have further expanded the design space for resource-efficient quantum computation~\cite{PhysRevA.111.012611,10682739}. Unlike static circuits, dynamic circuits enable adaptive behavior in which measurement outcomes determine subsequent gate executions. These measurement-driven feed-forward mechanisms can replace costly multi-qubit unitary subcircuits~\cite{10546695}. %10137250
Moreover, the availability of clean ancillas enables economical uncomputation through intermediate measurements, thereby reducing overhead~\cite{PhysRevA.87.022328,Gidney2018halvingcostof}.

%Decompositions of large Toffoli gates into Clifford+T networks typically involve auxiliary qubits, and the resulting T-count depends strongly on the availability and type of ancillas. In this context, relative-phase Toffoli gates have been shown to be effective in reducing both T-count and ancilla requirements~\cite{PhysRevA.93.022311}. Motivated by these developments, this work investigates the role of relative-phase Toffoli gates in the efficient realization of large Toffoli operations. Specifically, we study how 3- and 4-input relative-phase Toffoli gates can be exploited, in the presence of a single clean ancilla, to improve T-resource bounds when combined with conditionally clean ancillas and dynamic-circuit-based uncomputation.

%Decompositions of large Toffoli gates into Clifford+T networks depend strongly on the availability and type of ancillas. 
In decomposing large Toffoli gates, relative-phase Toffoli gates have been shown to reduce both T-count and ancilla requirements~\cite{PhysRevA.93.022311}. Motivated by these results, we study the use of 3- and 4-input relative-phase Toffoli gates with a single clean ancilla, combined with conditionally clean ancillas and dynamic uncomputation, to improve T-resource bounds. Although more T-depth-optimized constructions exist~\cite{PhysRevA.87.042302,10.1145/3649813}, they typically incur higher CX overhead. We therefore adopt simpler realizations~\cite{PhysRevA.93.022311} with constant CX cost to enable fair comparison and practical relevance for NISQ devices.
The main contributions of this work are summarized as follows:
\begin{enumerate}
\item We derive resource bounds for large Toffoli networks using 3- and 4-input relative-phase Toffoli gates with one clean and conditionally clean ancillas.
      
%We derive resource bounds for 3- and 4-input relative-phase Toffoli gates in the implementation of large Toffoli networks using conditionally clean ancillas and a single clean ancilla.
      
\item We show that 4-input relative-phase Toffoli gates reduce T-Depth by up to $2\lfloor (n-3)/6 \rfloor$ gate layers.
%We show that using 4-input relative-phase Toffoli gates reduces T-depth by up to $2\lfloor (n-3)/6 \rfloor$ times that of a single gate. 
%We show that introducing 4-input relative-phase Toffoli gates in the realization of $(n+1)$-input Toffoli gates yields a T-depth reduction of up to $2\lfloor (n-3)/6 \rfloor$ times the T-depth of a single 4-input Toffoli gate.
\item We incorporate dynamic-circuit-based uncomputation %of clean ancillas 
into the resource analysis. % for both 3- and 4-input relative-phase Toffoli constructions.
\item We experimentally evaluate the proposed constructions. %3- and 4-input relative-phase Toffoli approaches. 
%We present an experimental evaluation of the proposed approaches based on 3- and 4-input relative-phase Toffoli gates.
     
\end{enumerate}

\begin{comment}

\begin{figure*}[t]
    \centering
    \begin{subfigure}{0.10\linewidth}
        \centering
        \includegraphics[width=0.95\linewidth]{figures/RTOF1.pdf}
        \vspace{-0.28cm}
        \caption{}
        \label{fig:rtof1}
    \end{subfigure}
    \hspace{-0.5cm}
    \begin{subfigure}{0.37\linewidth}
        \centering
        \includegraphics[width=0.95\linewidth]{figures/RTOF1_decomp.pdf}
        \caption{}
        \label{fig:rtof1-decomp}
    \end{subfigure}
    %\hspace{-0.8cm}
    \begin{subfigure}{0.50\linewidth}
        \centering
        \includegraphics[width=0.95\linewidth]{figures/RTOF4.pdf}
        \vspace{0.09cm}
        \caption{}
        \label{fig:rtof4}
    \end{subfigure}

    \caption{Relative-phase Toffoli gate implementations: (a) the CC(iX) gate, (b) its Clifford+T decomposition, and (c) the Clifford+T realization of the C$^3$(iX) gate.}
    \label{fig:rtof}
\end{figure*}
\end{comment}

%The remainder of this paper is organized as follows. Section II provides the necessary background on quantum circuits, Toffoli operations, and related decomposition techniques. Section III presents our analysis of resource bounds for decomposing large Toffoli gates using one clean ancilla qubit and 3- and 4-input relative-phase Toffoli gates. Section IV reports experimental results comparing the use of 3- and 4-input relative-phase Toffoli gates. Finally, Section V concludes the paper.

The remainder of this paper is organized as follows. Section~II reviews background concepts and related decomposition techniques. Section~III presents the resource analysis for large Toffoli decompositions using one clean ancilla and 3- and 4-input relative-phase Toffoli gates. Section~IV reports experimental results, and Section~V concludes the paper.
\section{Background}
\subsection{Quantum Computation and Clifford+T Gate Set}
Quantum computation is based on the manipulation of quantum bits (qubits), which are represented by unit vectors in a two-dimensional complex Hilbert space $\mathcal{H}$. The computational basis states are
\begin{align}
\ket{0} =
\begin{pmatrix}
1 \\
0
\end{pmatrix}, \quad
\ket{1} =
\begin{pmatrix}
0 \\
1
\end{pmatrix},
\end{align}
and an arbitrary single-qubit state is given by $\ket{\psi}=\alpha_0\ket{0}+\alpha_1\ket{1}$, where $\alpha_0,\alpha_1\in\mathbb{C}$ and $|\alpha_0|^2+|\alpha_1|^2=1$. 
An $n$-qubit system is described by a superposition of computational basis states and evolves under unitary operations (quantum gates) arranged in a quantum circuit.

A widely used universal gate set is the Clifford+T basis. The Clifford group is generated by the Hadamard (H), phase (S), and controlled-NOT (CX) gates, while universality is achieved by supplementing these with the non-Clifford T gate. In fault-tolerant architectures, Clifford gates are relatively inexpensive, whereas T gates require costly magic-state distillation and injection. Consequently, non-Clifford operations dominate the cost of fault-tolerant circuits. This cost is commonly quantified using the \emph{T-count}, which measures the total number of T gates, and the \emph{T-depth}, which denotes the number of sequential layers of T gates.

\subsection{Dynamic Quantum Circuits}% and Measurement-Based Uncomputation}
Dynamic quantum circuits allows intermediate measurements, qubit resets, and classically controlled operations during circuit execution. Unlike static circuits, in which all gates are fixed in advance, dynamic circuits enable adaptive behavior where subsequent operations depend on measurement outcomes. 
This capability is particularly valuable in fault-tolerant and resource-constrained settings, where measurement-based feedback can be used to reduce circuit depth and ancilla overhead.

In many quantum algorithms, ancillary qubits are introduced to facilitate intermediate computations and must be uncomputed before being reused. Traditional uncomputation relies on reversing previously applied unitary operations, which often incurs substantial additional cost. In contrast, measurement-based uncomputation~\cite{PhysRevA.111.012611,10682739} exploits intermediate measurements on ancilla qubits, followed by classically conditioned corrections, to erase unwanted entanglement more efficiently. By replacing coherent uncomputation circuits with measurement-conditioned operations, this approach can significantly reduce the number of required non-Clifford gates.

\subsection{Multi-Controlled Toffoli Gates}

The Toffoli gate (CCX) is a three-qubit reversible gate that flips its target qubit $t$ only when both control qubits: $c_1$ and  $c_2$ are in the state $\ket{1}$. More generally, its $n$-control extension, denoted by C$^{n}$X, is defined as
\begin{align}\label{eq:CnX}
\text{C}^{n}\text{X}\ket{c_1\cdots c_n}\ket{t}
=
\ket{c_1\cdots c_n}\ket{t \oplus (c_1 \wedge \cdots \wedge c_n)}.
\end{align}

Such gates enable the implementation of complex Boolean functions and are fundamental components in many quantum algorithms. 
%However, direct physical implementations of $C^{n}X$ gates are typically unavailable, and they must therefore be decomposed into sequences of elementary gates. and multi-controlled Toffoli 
In the Clifford+T framework, Toffoli gates are non-Clifford operations and require decomposition into Clifford and T gates. Standard constructions often lead to high T-count and T-depth, making large Toffoli gates expensive in fault-tolerant settings.

As reported in~\cite{6516700}, the Clifford+T realization of the three-qubit Toffoli (CCX) gate requires 7 CX gates and 7 T gates with a T-depth of 3.
%, as illustrated in~\cref{fig:tof}. 
A T-depth of 1 can be achieved using the construction of~\cite{PhysRevA.87.042302}, at the cost of four ancilla qubits.
Various approaches have been proposed to optimize the realization of C$^{n}$X gates, including ancilla-assisted constructions~\cite{PhysRevA.111.052611,10.1007/978-3-031-08760-8_16,Khattar2025riseofconditionally,nie2024quantum}, relative-phase implementations~\cite{PhysRevA.93.022311}, and measurement-based techniques~\cite{PhysRevA.111.012611,10682739}. 
%In this work, we build upon these ideas to further reduce the resource cost of large Toffoli gates using relative-phase gates and dynamic circuit techniques.
In this work, we build upon these ideas to reduce the resource cost of large Toffoli gates using relative-phase gates and dynamic circuit techniques, while adopting simpler realizations with constant CX overhead to ensure fair comparison and practical relevance for NISQ devices.
\section{One-Clean Ancilla Decomposition}
\begin{figure}[t]
    \centering

    \begin{subfigure}{0.19\linewidth}
        \centering
        
        \includegraphics[width=0.95\linewidth]{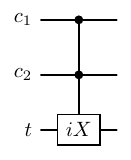}
        \vspace{-0.28cm}
        \caption{}
        \label{fig:rtof1}
        
    \end{subfigure}
    \hspace{-0.4cm}
    \begin{subfigure}{0.71\linewidth}
        \centering
        \includegraphics[width=0.95\linewidth]{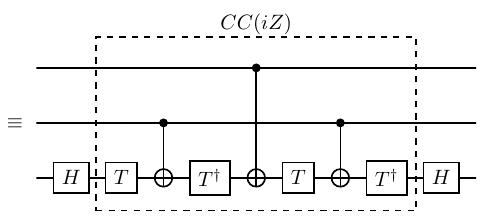}
        \caption{}
        \label{fig:rtof1-decomp}
    \end{subfigure}

    \begin{subfigure}{0.9\linewidth}
        \centering
        \includegraphics[width=0.95\linewidth]{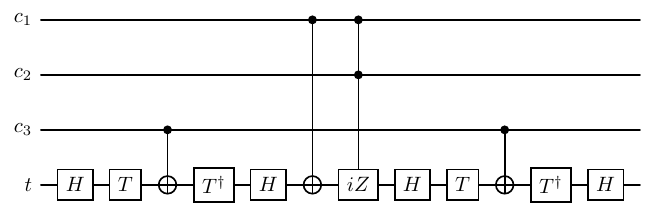}
        \caption{}
        \label{fig:rtof4}
    \end{subfigure}

    \caption{Relative-phase Toffoli gate implementations: (a) the CC(iX) gate, (b) its Clifford+T decomposition, and (c) the Clifford+T realization of the C$^3$(iX) gate.}
    \label{fig:rtof}
\end{figure}
\subsection{Clifford+T Optimization with CCiX and C3iX}
\begin{figure*}[t]
    \centering

    \begin{subfigure}{0.11\linewidth}
        \centering
        \includegraphics[width=0.73\linewidth]{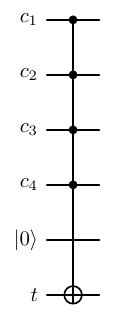}
        %\vspace{-0.28cm}
        \caption{}
        \label{fig:tof5}
        
    \end{subfigure}
    \hspace{-0.2cm}
    \begin{subfigure}{0.371\linewidth}
        \centering
        \includegraphics[width=0.95\linewidth]{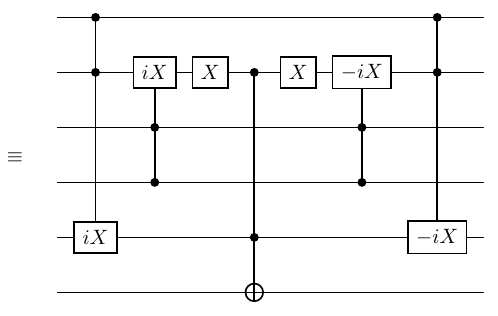}
        \caption{}
        \label{fig:tof5-decomp-1}
    \end{subfigure}
     \hspace{-0.1cm}
    \begin{subfigure}{0.193\linewidth}
        \centering
        \includegraphics[width=0.95\linewidth]{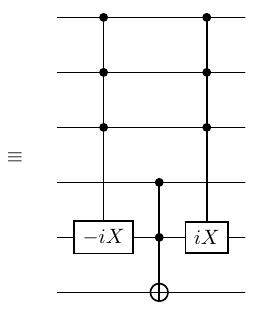}
        \caption{}
        \label{fig:tof5-decomp-2}
    \end{subfigure}

    \caption{(a) The realization of a 5-input Toffoli (C$^4$X) gate with one clean ancilla, implemented through two equivalent decompositions: (b) with four CC(iX) gates, and (c) with two C$^3$(iX) gates.}
    \label{fig:tof5-decomp}
\end{figure*}

\begin{figure}[t]
    \centering
    \includegraphics[width=0.75\linewidth]{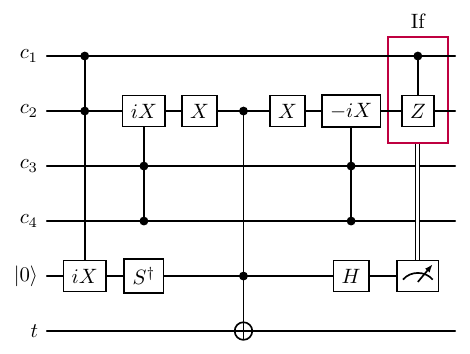}
    \caption{A dynamic realization of the C$^4$X gate employing three CC(iX) gates and a conditional CZ operation triggered by the ancilla measurement.}
    \label{fig:tof5-dynamic-1}
\end{figure}

Relative-phase Toffoli gates have proven advantageous
for realizing higher-order Toffoli gates C$^{n}$X ($n>2$), as demonstrated in~\cite{PhysRevA.93.022311}. \Cref{fig:rtof} illustrates the Clifford+T implementations of the
3-input CC(iX) and 4-input C$^3$(iX) relative-phase Toffoli gates. The $C^3$(iX) construction exactly doubles the Clifford+T resources
required for CC(iX), that is,
\begin{align}\label{eq:cost}
    \mathcal{C}_\mathrm{ost}\big(\text{C$^{3}$(iX)\big)} = 2~\mathcal{C}_\mathrm{ost}\big(\text{CC(iX)}\big).
\end{align} 
Nevertheless, the use of C$^3$(iX) gates is beneficial for minimizing
the number of ancilla qubits.

Using conditionally clean ancilla qubits~\cite{Khattar2025riseofconditionally,nie2024quantum}, a C$^{4}$X gate can be implemented as a network of CC(iX) gates with a
single clean ancilla, as shown in~\cref{fig:tof5-decomp-1}. Since CC(iX) requires 3 CXs, 4 T gates, and has T-depth 4 (see~\cref{fig:rtof1-decomp}),
the resulting construction using four CC(iX) gates and one CCX gate has the total cost  
\begin{align}\label{eq:tof5-decomp-1}
4 \times (3~\text{CX},\, 4~\text{T-count},\, 4~\text{T-depth}) \;\;\;\;\;\;\nonumber\\\; +\;(7~\text{CX},\, 7~\text{T-count},\, 3~\text{T-depth}).    
\end{align}%+ 2~\text{X}
The two additional X gates enable the control qubit $c_2$ to act as a conditionally clean ancilla.

\begin{figure}[t]
    \centering
    \includegraphics[width=0.75\linewidth]{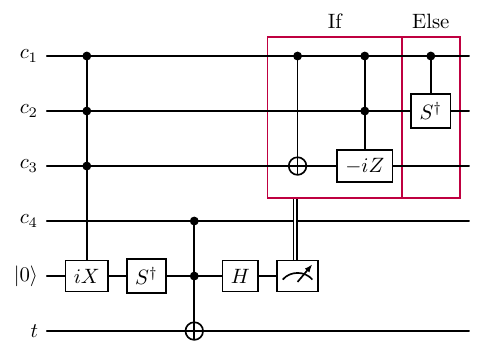}
    \caption{A dynamic implementation of the C$^4$X gate using one C$^3$(iX) gate together with measurement-conditioned operations: CX$\cdot$CC(iZ) when the ancilla outcome is 1, and CS$^\dagger$ when it is 0.}
    \label{fig:tof5-dynamic-2}
\end{figure}

An alternative realization based on a single clean ancilla was proposed in~\cite{PhysRevA.93.022311}, using C$^{3}$(iX) gates. As illustrated in~\cref{fig:tof5-decomp-2}, this construction employs two C$^{3}$(iX) gates and yields the cost 
\begin{align}\label{eq:tof5-decomp-2}
2 \times (6~\text{CX},\, 8~\text{T-count},\, 8~\text{T-depth}) \;\;\;\;\;\;\;\;\;\;\nonumber\\\; +\;(7~\text{CX},\, 7~\text{T-count},\, 3~\text{T-depth}).
\end{align}
Both approaches in~\cref{eq:tof5-decomp-1,eq:tof5-decomp-2} therefore incur identical overall costs. However, the C$^{3}$(iX)-based construction avoids
the additional X gates, as shown in~\cref{fig:tof5-decomp}.

\begin{figure*}[t]
    \centering

         \includegraphics[width=1.0\linewidth]{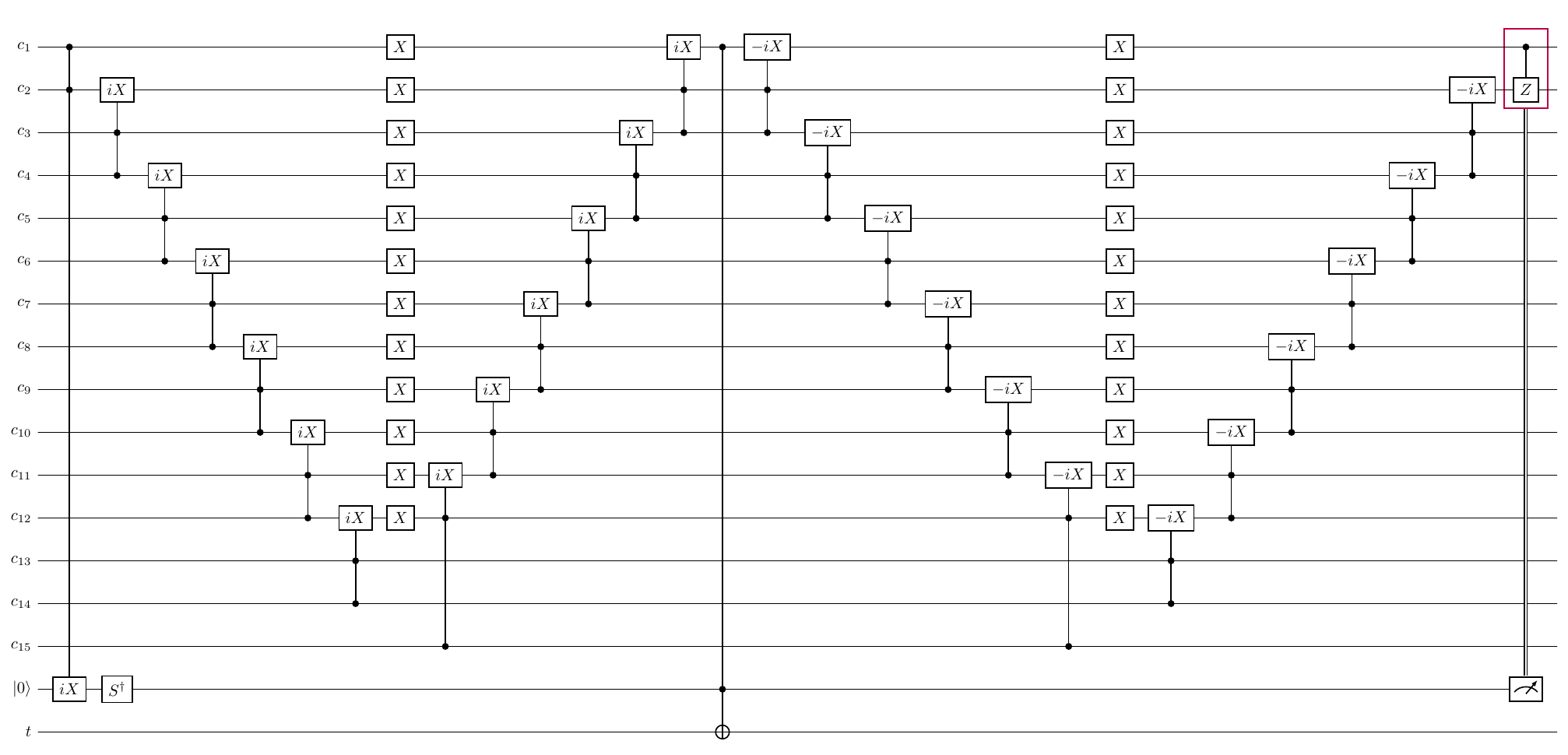}
    
    \caption{A dynamic realization of the C$^{15}$X gate with one clean ancilla, using a network of 25 CC(iX) gates, one CCX gate, and a conditional CZ operation triggered by ancilla measurement, derived from~\cite{Khattar2025riseofconditionally}.}
    \label{fig:tof16-decomp-1}
\end{figure*}

Further reductions can be achieved using dynamic circuits. According to~\cite{PhysRevA.87.022328}, a fault-tolerant CCX gate can be implemented with a clean ancilla using only four T gates. Applying this approach to the C$^4$X decomposition in~\cref{fig:tof5-dynamic-1}, and replacing the final CC(iX) gate by an
ancilla phase correction ($S^\dagger$), a Hadamard-basis measurement, and measurement-conditioned phase restoration, yields
\begin{align}\label{eq:tof5-dynamic-decomp-1}
3 \times (3~\text{CX},\, 4~\text{T-count},\, 4~\text{T-depth}) \;\;\;\;\;\;\;\;\;\;\;\;\;\;\;\;\;\;\;\;\nonumber\\\; +\;(7~\text{CX},\, 7~\text{T-count},\, 3~\text{T-depth}) \; +\; [1~\text{CX}].    
\end{align}
Here, $[1~\text{CX}]$ denotes an optional CX applied only when the Hadamard-basis measurement outcome of the clean ancilla is one, i.e. $M_H=1$. Compared with~\cref{eq:tof5-decomp-1,eq:tof5-decomp-2}, this construction reduces the T-count and T-depth by four and the CX count by two.

A similar dynamic approach can be applied to the C$^3$(iX)-based network in~\cref{fig:tof5-dynamic-2}, leading to
\begin{align}\label{eq:tof5-dynamic-decomp-2}
(6~\text{CX},\, 8~\text{T-count},\, 8~\text{T-depth}) \;\;\;\;\;\;\;\;\;\;\;\;\;\;\;\;\;\;\;\;\;\;\;\;\;\;\;\;\;\;\;\;\nonumber\\\; +\;(7~\text{CX},\, 7~\text{T-count},\, 3~\text{T-depth}) \;\;\;\;\;\;\;\;\;\;\;\;\;\;\;\;\;\;\;\;\;\;\nonumber\\
+\;\begin{cases}
(4~\text{CX},\, 4~\text{T-count},\, 4~\text{T-depth}),& M_H = 1\\[4pt]
(2~\text{CX},\, 3~\text{T-count},\, 2~\text{T-depth}),& M_H = 0.
\end{cases}
\end{align}
It replaces the final C$^3$(iX) gate by an ancilla phase correction ($S^\dagger$), a Hadamard-basis measurement, and measurement-conditioned phase restoration of the form $\text{CX}\cdot\text{CC(iZ)}$ for $M_H = 1$ and $CS^\dagger$ for $M_H = 0$.

When $M_H = 1$, both dynamic constructions in~\cref{eq:tof5-dynamic-decomp-1,eq:tof5-dynamic-decomp-2} have comparable costs (ignoring X gates). When $M_H = 0$, however, the construction in~\cref{fig:tof5-dynamic-2} is more efficient, requiring one fewer CX and T gate and reducing the T-depth by two.

%Replacing the final C$^3$(iX) gate by an ancilla phase correction ($S^\dagger$), a Hadamard-basis measurement, and measurement-conditioned phase restoration of the form CX$\cdot$CC(iZ) for $M_H = 1$.  
%The latter case ($M_H = 0$) corresponds to the more economical use of the  $CS^\dagger$ gate~\cite{PhysRevA.52.3457}. When $M_H(a)=1$, both dynamic constructions
%in~\cref{eq:tof5-dynamic-decomp-1,eq:tof5-dynamic-decomp-2} have comparable costs (ignoring the X gates). When $M_H=0$, however, the construction of~\cref{fig:tof5-dynamic-2} is more efficient, requiring one fewer CX and T gate and reducing the T-depth by two.

\subsection{Depth Minimization with C3iX over CCiX}
According to~\cite{Khattar2025riseofconditionally}, the use of conditionally clean ancilla enables the realization of a C$^n$X gate using $2n-3$ CCX gates when one clean ancilla is used. The use of CC(iX) gates in the network enables the realization to be obtained using $2n-4$ CC(iX) gates and one CCX gate. Replacing the final CC(iX) gate by an ancilla phase correction ($S^\dagger$), a Hadamard-basis measurement, and measurement-conditioned phase restoration further improves the resource bounds of the one-clean-ancilla-based realization:
\begin{align}\label{eq:tofn-dynamic-decomp-1}
(2n-5) \times (3~\text{CX},\, 4~\text{T-count},\, 4~\text{T-depth}) \;\;\;\;\;\;\nonumber\\
 +\;(7~\text{CX},\, 7~\text{T-count},\, 3~\text{T-depth}) \; +\; [1~\text{CX}].
\end{align}
As an illustration,~\cref{fig:tof16-decomp-1} shows the realization of a C$^{15}$X gate as a network of 25 CC(iX) gates.

\begin{figure}[t]
    \centering

         \includegraphics[width=1.0\linewidth]{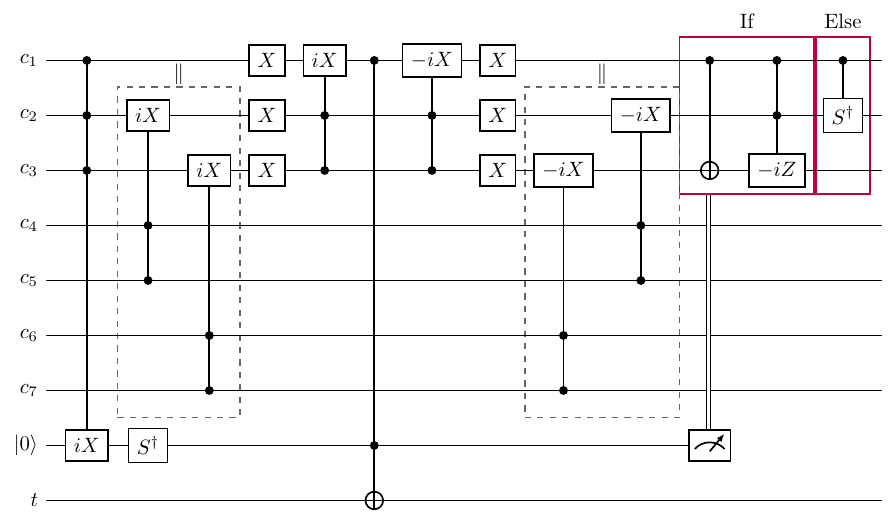}
    
    \caption{A dynamic realization of the C$^{7}$X gate with reduced T-depth and one clean ancilla. The use of the C$^3$(iX) gate enables enables two successive parallel pairs of CC(iX) gates acting on the same set of qubits. The final C$^3$(iX) gate is replaced by measurement-conditioned operations: $\text{CX}\cdot\text{CC(iZ)}$ when the ancilla outcome is 1, and $CS^\dagger$ when it is 0.}
    \label{fig:tof7-decomp-1}
\end{figure}

\begin{figure*}[t!]
    \centering

         \includegraphics[width=0.8\linewidth]{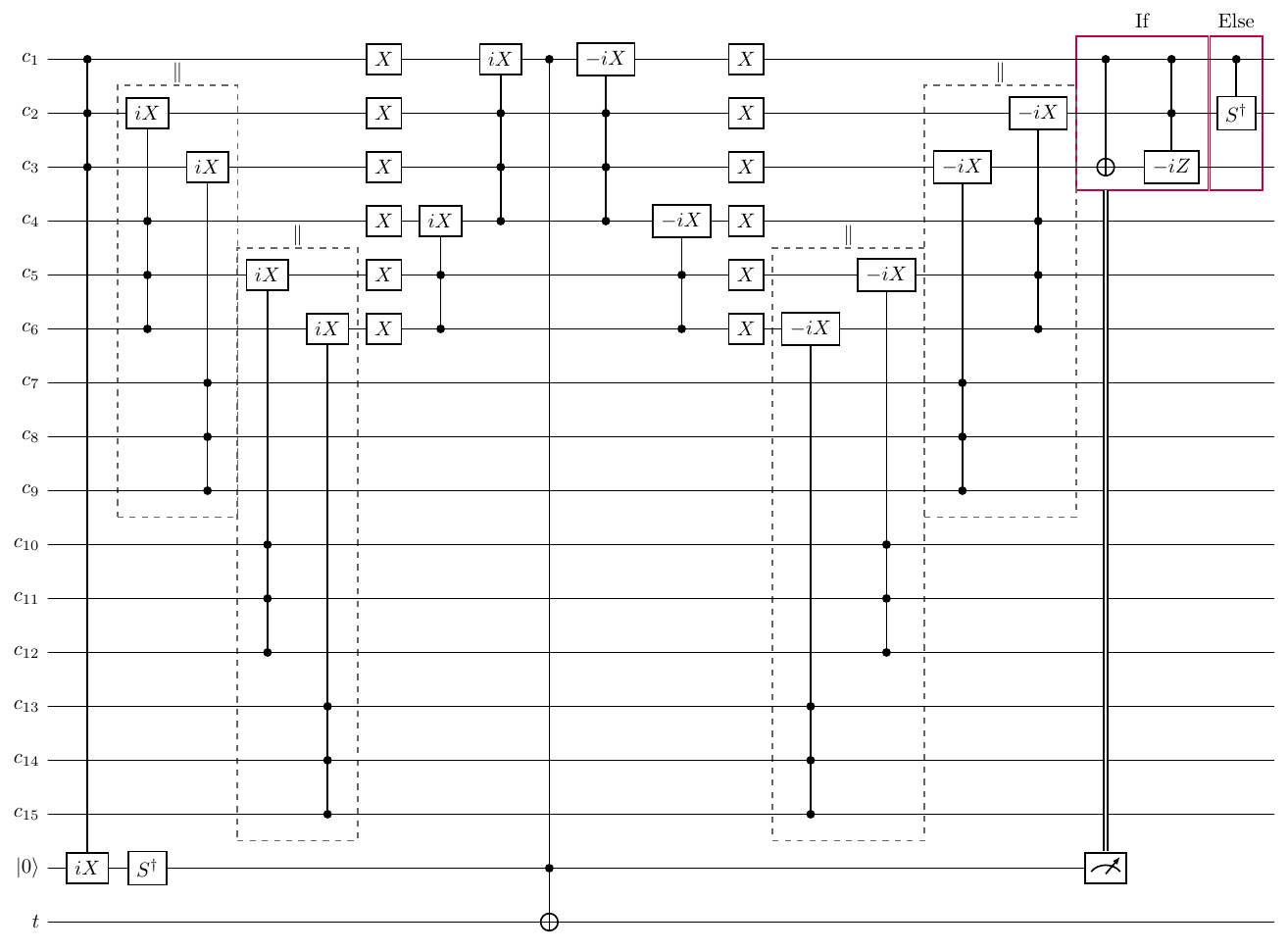}
    
    \caption{A dynamic realization of the C$^{15}$X gate with one clean ancilla. The use of the C$^3$(iX) gate enables four parallel pairs of C$^3$(iX) gates. The final C$^3$(iX) gate is replaced by measurement-conditioned operations: $\text{CX}\cdot\text{CC(iZ)}$ when the ancilla outcome is 1, and $CS^\dagger$ when it is 0.}
    \label{fig:tof16-decomp-2}
\end{figure*}

The decomposition using the C$^3$(iX) gate does not reduce the overall gate count, as it requires twice the resources of the CC(iX) gate (see~\cref{eq:cost}). However, its use increases the scope for parallel gate scheduling and thereby reduces both the T-depth and the overall circuit depth in realizing larger Toffoli operations. This improvement arises from the fact that C$^3$(iX) provides three conditionally clean ancilla, compared to two in the case of CC(iX). The benefits of such parallelism become evident when decomposing a C$^n$X gate for $n \geqslant 7$, as shown in~\cref{fig:tof7-decomp-1}.

By shifting a CC(iX) gate to serve as a measurement-conditioned phase correction CC(iZ), the static\footnote{Here, \emph{static} refers to circuit components that are not conditioned on classical measurement outcomes.} cost of the C$^3$(iX)-based dynamic decomposition is reduced by one CC(iX) gate compared to the CC(iX)-based dynamic decomposition. According to~\cref{eq:tofn-dynamic-decomp-1}, the realization of a C$^7$X gate requires 9 CC(iX) gates, whereas the use of C$^3$(iX) gates reduces this requirement to 6 CC(iX) gates (see~\cref{fig:tof7-decomp-1}). However, since the cost of a C$^3$(iX) gate is twice that of a CC(iX) gate (see~\cref{eq:cost}), the overall static cost of the C$3$(iX)-based dynamic decomposition corresponds to 8 CC(iX) gates. Nevertheless, due to the parallel execution of two successive pairs of CC(iX) gates (see~\cref{fig:tof7-decomp-1}), the T-depth is reduced by $2 \times \text{T-depth}(\text{CC(iX)})$. For $n \geqslant 7$, the realization %of a C$^n$X gate 
admits an additional T-depth reduction of
\begin{align}\label{eq:td}
2\,\mathbf{1}_{\{\,n \bmod 6 \in \{1,2\}\,\}} \times \text{T-depth}(\text{CC(iX)}),
\end{align}
i.e., two successive pairs of CC(iX) gates can be scheduled in parallel when $n = 6k+1$, $k \geqslant 1$.

For $n \geqslant 9$, the T-depth reduction can be further increased, since the realization of a C$^n$X gate enables the parallel scheduling of at least two successive pairs of C$^3$(iX) gates. As C$^3$(iX) is twice as costly as CC(iX), parallel execution of these pairs yields an additional factor of two improvement in T-depth. Since each pair of C$^3$(iX) gates operates on six distinct qubits, the maximum T-depth reduction for $n \geqslant 9$ is given by
\begin{align}\label{eq:td2}
2\Big\lfloor\frac{n-3}{6}\Big\rfloor \times \text{T-depth}(\text{C$^3$(iX)}).
\end{align}

Consequently, the C$^3$(iX)-based dynamic decomposition improves the resource bounds of the C$^n$X realization as
\begin{align}\label{eq:tofn-dynamic-decomp-2}
(2n-6) \times (3~\text{CX},\, 4~\text{T-count},\, 4~\text{T-depth}) \;\;\;\;\;\;\;\;\;\;\;\;\;\nonumber\\
+\; (7~\text{CX},\, 7~\text{T-count},\, 3~\text{T-depth}) \;\;\;\;\;\;\;\;\;\;\;\;\;\;\;\;\;\;\;\;\;\;\nonumber\\
+\;\begin{cases}
(4~\text{CX},\, 4~\text{T-count},\, 4~\text{T-depth}),& M_H = 1,\\[4pt]
(2~\text{CX},\, 3~\text{T-count},\, 2~\text{T-depth}),& M_H = 0,
\end{cases}
\nonumber\\
-\;\,2\Big\lfloor\frac{n-3}{6}\Big\rfloor \times \text{T-depth}(\text{C$^3$(iX)}) \;\;\;\;\;\;\;\;\;\;\;\;\;\;\;\;\;\;\;\;\;\;\nonumber\\
-\,2\,\mathbf{1}_{\{\,n \bmod 6 \in \{1,2\}\,\}} \times \text{T-depth}(\text{CC(iX)}).\;\;\;\;\;\;\;\;
\end{align}
Thus, compared to~\cref{eq:tofn-dynamic-decomp-1}, the introduction of C$^3$(iX) yields a more T-depth-efficient realization. As an illustration,~\cref{fig:tof16-decomp-2} shows the realization of a C$^{15}$X gate as a static network of 11 C$^3$(iX) gates and 2 CC(iX) gates, enabling four pairs of parallel C$^3$(iX) gate executions.

{ \setlength{\tabcolsep}{8pt}
\begin{table*}[t!]
  \caption{\small{Results for the dynamic realization of the Toffoli operation using one clean ancilla, comparing the method of~\cite{Khattar2025riseofconditionally} with the proposed optimization.}}
  \label{tab1}
%  \begin{minipage}{\textwidth}
   \begin{center}
     \begin{tabular}{r||rrr||rrr|rrr||rrr|rrr} \toprule 
 & \multicolumn{3}{c||}{Static~\cite{Khattar2025riseofconditionally}} & \multicolumn{3}{c|}{CC(iX)} & \multicolumn{3}{c||}{Impr.(\%)} & \multicolumn{3}{c|}{CC(iX)$\,+\,$C$^3$(iX)} & \multicolumn{3}{c}{Impr.(\%)}\\ \cmidrule(lr){2-4}\cmidrule(lr){5-10}\cmidrule(lr){11-16} 
$n$& $CX$ & $T_c$ & $T_d$ & $CX$ & $T_c$ & $T_d$ & $CX$ & $T_c$ & $T_d$ & $CX$ & $T_c$ & $T_d$& $CX$ & $T_c$ & $T_d$ \\ \midrule
4 & 19 & 23 & 19 & 17 & 19 & 15 & 10.53 & 17.39 & 21.05 & 17 & 19 & 15 & 10.53 & 17.39 & 21.05 \\
5 & 25 & 31 & 27 & 23 & 27 & 19 & 8.00 & 12.90 & 29.63 & 23 & 27 & 20 & 8.00 & 12.90 & 25.93 \\
6 & 31 & 39 & 35 & 29 & 35 & 27 & 6.45 & 10.26 & 22.86 & 29 & 35 & 28 & 6.45 & 10.26 & 20.00 \\
7 & 37 & 47 & 43 & 35 & 43 & 31 & 5.41 & 8.51 & 27.91 & 35 & 43 & 29 & 5.41 & 8.51 & 32.56 \\
8 & 43 & 55 & 51 & 41 & 51 & 39 & 4.65 & 7.27 & 23.53 & 41 & 51 & 33 & 4.65 & 7.27 & 35.29 \\
9 & 49 & 63 & 59 & 47 & 59 & 43 & 4.08 & 6.35 & 27.12 & 47 & 59 & 37 & 4.08 & 6.35 & 37.29 \\
10 & 55 & 71 & 67 & 53 & 67 & 51 & 3.64 & 5.63 & 23.88 & 53 & 67 & 43 & 3.64 & 5.63 & 35.82 \\
11 & 61 & 79 & 75 & 59 & 75 & 55 & 3.28 & 5.06 & 26.67 & 59 & 75 & 51 & 3.28 & 5.06 & 32.00 \\
12 & 67 & 87 & 83 & 65 & 83 & 63 & 2.99 & 4.60 & 24.10 & 65 & 83 & 53 & 2.99 & 4.60 & 36.14 \\
13 & 73 & 95 & 91 & 71 & 91 & 67 & 2.74 & 4.21 & 26.37 & 71 & 91 & 55 & 2.74 & 4.21 & 39.56 \\
14 & 79 & 103 & 99 & 77 & 99 & 75 & 2.53 & 3.88 & 24.24 & 77 & 99 & 59 & 2.53 & 3.88 & 40.40 \\
15 & 85 & 111 & 107 & 83 & 107 & 79 & 2.35 & 3.60 & 26.17 & 83 & 107 & 63 & 2.35 & 3.60 & 41.12 \\
16 & 91 & 119 & 115 & 89 & 115 & 87 & 2.20 & 3.36 & 24.35 & 89 & 115 & 67 & 2.20 & 3.36 & 41.74
 \\\bottomrule
%\multicolumn{16}{l}{CC(iX)$^{*}\xrightarrow{}$ Cost of the one-clean-ancilla dynamic realization using CC(iX) gates.}\\
%\multicolumn{16}{l}{CC(iX)$\,+\,$C$^3$(iX)$^{*}\xrightarrow{}$ Cost of the one-clean-ancilla dynamic realization using CC(iX) %and C$^3$(iX) gates.}\\
%\multicolumn{16}{l}{$n\xrightarrow{} \mathrm{C}^{n}X$ gate; $\mathrm{C}X \xrightarrow{}\mathrm{C}X$-Count; $T_c\xrightarrow{}T$-Count; $T_d \xrightarrow{}T$-Depth.}\\\bottomrule
      \end{tabular}
    \end{center}
\end{table*}
}
\section{Experimental Results}
For experimental evaluation, we consider C$^{n}$X gates with $4 \leqslant n \leqslant 16$. As a baseline, we adopt the one-clean-ancilla realization proposed in~\cite{Khattar2025riseofconditionally}, whose resource cost is given by
\begin{align}
  (2n-4) \times (3~\text{CX},\, 4~\text{T-count},\, 4~\text{T-depth}) \;\;\;\;\;\;\nonumber\\
 +\;(7~\text{CX},\, 7~\text{T-count},\, 3~\text{T-depth}). 
\end{align}

For fair comparison, we implement both the CC(iX)-based and the combined CC(iX)$\,+\,$C$^3$(iX)-based constructions and evaluate their resource costs in terms of CX-count ($CX$), T-count ($T_c$), and T-depth ($T_d$). Moreover, we consider the worst-case measurement-driven uncomputation cost, corresponding to the case $M_H=1$, in all evaluations.

The resulting performance metrics are summarized in~\cref{tab1}. 
%As observed from the table, while the CX-count and T-count remain identical, the combined CC(iX)$\,+\,$C$^3$(iX) construction consistently achieves greater T-depth reductions than the CC(iX)-based approach for implementing C$^{n}$X gates with $n \geqslant 7$, in accordance with the theoretical bounds derived in~\cref{eq:tofn-dynamic-decomp-1,eq:tofn-dynamic-decomp-2}.
As observed from the table, while the CX-count and T-count remain identical, the combined CC(iX)$\,+\,$C$^3$(iX) construction achieves greater T-depth reductions than the CC(iX)-based approach for C$^{n}$X gates with $n \geqslant 7$, in accordance with the theoretical bounds derived in~\cref{eq:tofn-dynamic-decomp-1,eq:tofn-dynamic-decomp-2}.

\section{Conclusion}
In this work, we studied the efficient realization of large Toffoli gates using relative-phase constructions, conditionally clean ancillas, and dynamic circuit techniques. Using 3- and 4-input relative-phase Toffoli gates with a single clean ancilla, we derived resource bounds and analyzed their impact on T-count and T-depth. Our results show that higher-input relative-phase Toffoli gates enable significant T-depth reductions through enhanced parallelism, while dynamic uncomputation and measurement-conditioned corrections further reduce overhead. Experimental results confirm consistent improvements over existing methods, highlighting the effectiveness of combining relative-phase and dynamic-circuit techniques for scalable quantum computation.

Future work will investigate constructions that further optimize T-depth by allowing controlled increases in CX overhead, thereby exploring the tradeoff between two-qubit gate cost and circuit latency.

%Future work may extend this framework to hardware-specific implementations and further investigate dynamic-circuit-based uncomputation and the efficient use of conditionally clean ancillas for larger multi-controlled operations. Such developments are expected to enable more resource-efficient quantum algorithms and accelerate progress toward practical fault-tolerant quantum computing.   

%\section{Acknowledgments}
%\thanks{This work was partly funded by the Federal Ministry of Research, Technology and Space (BMFTR; formerly BMBF) through the EASEPROFIT project (grant no. 16KIS2127) and by the German Research Foundation (DFG) through the CONAD-QC project (grant no. 559888852). The research is conducted within the scope of the DFG Priority Programme 2514 (SPP 2514).}

\bibliographystyle{IEEEtran}
\bibliography{Toff}

\end{document}